\documentclass[twocolumn,twoside,slac_two]{revtex4}
\usepackage{graphicx}
\usepackage{fancyhdr}
\usepackage{epsfig}
\usepackage{epsf}
\newcommand{\BR}{{\cal B}}
\newcommand{\psp}{\psi(2S)}
\newcommand{\jpsi}{J/\psi}
\newcommand{\pipi}{\pi^+\pi^-}
\newcommand{\kskp}{K^0_S K^+ \pi^- + c.c.}
\newcommand{\ks}{K^0_S}
\newcommand{\kk}{K^+K^-}
\newcommand{\ppb}{p\overline{p}}
\newcommand{\ld}{\Lambda}
\newcommand{\ldb}{\bar\Lambda}
\newcommand{\ar}{\rightarrow}
\newcommand{\llb}{\Lambda\bar{\Lambda}}
\newcommand{\llp}{\llb\pi^0}
\newcommand{\lle}{\llb\eta}
\newcommand{\splb}{\Sigma^+\pi^-\bar{\Lambda}}
\newcommand{\sbpl}{\bar{\Sigma}^-\pi^+\ld}

\newcommand{\lmb}{\bar{\Lambda}}
\newcommand{\GG}{\gamma\gamma}
\newcommand{\bcl}{\begin{center}}
\newcommand{\ecl}{\end{center}}
\newcommand{\bbt}{\bibitem}
\newcommand{\beq}{\begin{equation}}
\newcommand{\eeq}{\end{equation}}
\newcommand{\beqr}{\begin{eqnarray}}
\newcommand{\eeqr}{\end{eqnarray}}

\begin{document}

\title{Recent Results From BESII}


\author{Rong-Gang Ping (for the BES Collaboration)}
\affiliation{Institute of High
Energy Physics, Chinese Academy of Sciences, \\
P.O. Box 918-1, Beijing 100049, China}%

\begin{abstract}
We present recent results from the BES experiment on the observation
of the $Y(2175)$ in $J/\psi\to \phi f_0(980) \eta$, and
$\eta(2225)$ in $J/\psi\to \gamma \phi \phi$, and $X(1440)$ in $J/\psi$
hadronic decays, together with the new observation of $\psi(2S)$ radiative decays and hadronic decays into
$n\ks\ldb+c.c.,~\ld\ldb\pi^0,~\ld\ldb\eta$. The effort to search for $J/\psi$ decays into $\gamma\gamma$ and invisible decays are also reported.
 \vspace{1pc}
\end{abstract}

\maketitle

\thispagestyle{fancy}


\section{Introduction}

The analyses reported in this talk were performed using either a
sample of $58 \times 10^{6}$ $J/\psi$ events or a sample of $14
\times 10^{6}$ $\psi(2S)$ events collected with the upgraded Beijing
Spectrometer (BESII) detector~\cite{BESII} at the Beijing
Electron-Positron Collider (BEPC).
\section{Light hadron spectroscopy}
\subsection{The $Y(2175)$ in $J/\psi\to \phi f_0(980)
\eta$~\cite{bes2175}}

A new structure, denoted as $Y(2175)$ and with mass $m=2.175\pm
0.010\pm 0.015$~GeV/$c^2$ and width $\Gamma=58\pm 16\pm
20$~MeV/$c^2$, was observed by the BaBar experiment in the
$e^+e^-\to\gamma_{ISR}\phi f_0(980)$ initial-state radiation
process~\cite{babary21752006,babary21752007}. This observation
stimulated some theoretical speculation that this $J^{PC}=1^{--}$
state may be an $s$-quark version of the $Y(4260)$ since both of
them are produced in $e^+e^-$ annihilation and exhibit similar decay
patterns~\cite{babar4260,belle4260}.

\begin{figure}[htbp]
  \centering
\includegraphics[width=0.40\textwidth]{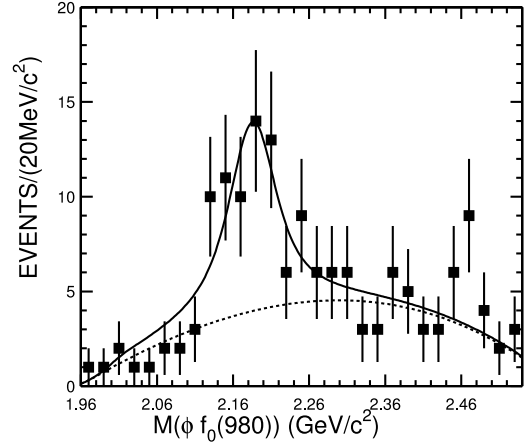}
\caption{The $\phi f_0(980)$ invariant mass distribution of the data
(points with error bars) and the fit (solid curve) with a
Breit-Wigner function and polynomial background; the dashed curve
indicates the background function.} \label{draft-fit}
\end{figure}

Here we report the observation of the $Y(2175)$ in the decays of
$J/\psi\to \eta \phi f_0(980)$, with $\eta\to \gamma\gamma$, $\phi
\to K^+K^-$, $f_0(980)\to\pi^+\pi^-$. A four-constraint
energy-momentum conservation kinematic fit is performed to the $K^+
K^-\pi^+\pi^-\gamma\gamma$ hypothesis for the selected four charged
tracks and two photons. $\eta\to\gamma\gamma$ candidates are defined
as $\gamma$-pairs with $|M_{\gamma\gamma}-0.547|<0.037$~GeV/$c^2$, a
$\phi$ signal is defined as $|m_{K^+K^-}-1.02|<0.019$~GeV/$c^2$, and
in the $\pi^+\pi^-$ invariant mass spectrum, candidate $f_0(980)$
mesons are defined by $|m_{\pi^+\pi^-}-0.980|<0.060$~GeV/$c^2$. The
$\phi f_0(980)$ invariant mass spectrum for the selected events is
shown in Fig.~\ref{draft-fit}, where a clear enhancement is seen
around 2.18~GeV/$c^2$. Fit with a Breit-Wigner and a polynomial
background yields $52\pm12$ signal events and the statistical
significance is found to be $5.5\sigma$ for the signal. The mass of
the structure is determined to be $M=2.186\pm 0.010~(stat)\pm
0.006~(syst)$~GeV/$c^2$, the width is $\Gamma=0.065\pm
0.023~(stat)\pm 0.017~(syst)$~GeV/$c^2$, and the product branching
ratio is $\BR(J/\psi \to \eta Y(2175))\cdot \BR(Y(2175)\to \phi
f_0(980))\cdot \BR(f_0(980)\to\pi^+\pi^-)=(3.23\pm 0.75~(stat)\pm
0.73~(syst))\times 10^{-4}$. The mass and width are consistent with
BaBar's results.

\subsection{The $\eta(2225)$ in $J/\psi\to \gamma
\phi\phi$~\cite{bes2225}}

Structures in the $\phi\phi$ invariant-mass spectrum have been
observed by several experiments both in the reaction
$\pi^{-}p\to\phi\phi n$~\cite{pip} and in radiative $J/\psi$
decays~\cite{mk3,dm21,dm22}. The $\eta(2225)$ was first observed by
the MARK-III collaboration in $J/\psi$ radiative decays $J/\psi \to
\gamma\phi\phi$. A fit to the  $\phi\phi$ invariant-mass spectrum
gave a mass of 2.22~GeV/$c^2$ and a width of
150~MeV/$c^2$~\cite{mk3}. An angular analysis of the structure found
it to be consistent with a $0^{-+}$ assignment. It was subsequently
observed by the DM2 collaboration, also in $J/\psi \to \gamma \phi
\phi$ decays~\cite{dm21,dm22}.

\begin{figure}[htbp]
  \centering
\includegraphics[width=0.40\textwidth]{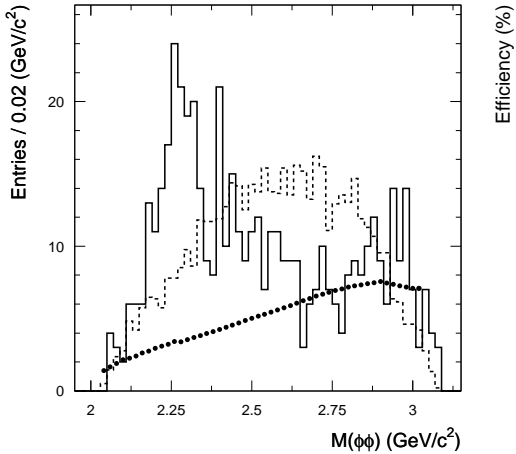}
\caption{The $K^+K^-K^0_SK^0_L$ invariant mass distribution for
$J/\psi \to \gamma \phi\phi$ candidate events. The dashed histogram
is the phase space invariant mass distribution, and the dotted curve
indicates how the acceptance varies with the $\phi\phi$ invariant
mass.}   \label{dalitz}
\end{figure}

We present results from a high statistics study of $J/\psi \to
\gamma \phi \phi$ in the $\gamma K^+ K^- K^0_S K^0_L$ final state,
with the $K^0_L$ missing and reconstructed with a one-constraint
kinematic fit. After kinematic fit, we require both the $K^+K^-$ and
$K^0_SK^0_L$ invariant masses lie within the $\phi$ mass region
($|M(K^+K^-)-m_{\phi}|<12.5$~MeV/$c^2$ and
$|M(K^0_SK^0_L)-m_{\phi}|<25$~MeV/$c^2$). The $\phi\phi$ invariant
mass distribution is shown in Fig.~\ref{dalitz}. There are a total
of 508 events with a prominent structure around 2.24~GeV/$c^2$.

A partial wave analysis of the events with $M(\phi\phi)<$
2.7~GeV/$c^2$ was performed. The two-body decay amplitudes in the
sequential decay process $J/\psi \to \gamma X, X\to \phi\phi$,
$\phi\to K^+ K^- $ and $\phi\to K^0_S K^0_L$ are constructed using
the covariant helicity coupling amplitude method. The intermediate
resonance $X$ is described with the normal Breit-Wigner propagator
$BW = 1/(M^2-s-i M\Gamma)$, where $s$ is the $\phi\phi$ invariant
mass-squared and $M$ and $\Gamma$ are the resonance's mass and
width. When $J/\psi \to \gamma X$, $X \to \phi \phi$ is fitted with
both the $\phi\phi$ and $\gamma X$ systems in a $P$-wave, which
corresponds to a pseudoscalar $X$ state, the fit gives $196\pm 19$
events with mass $M
=2.24^{+0.03}_{-0.02}{}^{+0.03}_{-0.02}$~GeV/$c^2$, width $\Gamma
=0.19\pm0.03^{+0.04}_{-0.06}$~GeV/$c^2$, and a statistical
significance larger than $10\sigma$, and a product branching
fraction of: $\BR(J/\psi \to \gamma \eta(2225))\cdot
\BR(\eta(2225)\to \phi\phi)=(4.4\pm 0.4\pm 0.8)\times 10^{-4}$.

The presence of a signal around 2.24~GeV/$c^2$ and its pseudoscalar
character are confirmed, and the mass, width, and branching fraction
are in good agreement with previous experiments.

\subsection{The $X(1440)$ in $J/\psi$ hadronic decays~\cite{bes1440}}

A pseudoscalar gluonium candidate, the so-called $E/\iota(1440)$,
was observed in $p\bar{p}$ annihilation in 1967~\cite{baillon67} and
in $J/\psi$ radiative decays in the
1980's~\cite{scharre80,edwards82e,augustin90}. The study of the
decays $J/\psi \rightarrow$ \{$\omega$, $\phi$\}$K\bar{K}\pi$ is a
useful tool in the investigation of quark and possible gluonium
content of the states around 1.44~GeV/$c^{2}$.  Here we investigate
the possible structure in the $K\bar{K}\pi$ final state in $J/\psi$
hadronic decays at around $1.44$~GeV/$c^{2}$.

In this analysis, $\omega$ mesons are observed in the $\omega
\rightarrow \pi^{+}\pi^{-}\pi^{0}$ decay, $\phi$ mesons in the $\phi
\rightarrow K^{+}K^{-}$ decay, and other mesons are detected in the
decays: $K^{0}_{S} \rightarrow \pi^{+}\pi^{-}$, $\pi^0 \rightarrow
\gamma \gamma$. $K\bar{K}\pi$ could be $K^{0}_{S}K^{\pm} \pi^{\mp}$
or $K^{+}K^{-}\pi^{0}$.

\begin{figure*}[htbp]
  \centering
\includegraphics[width=0.43\textwidth]{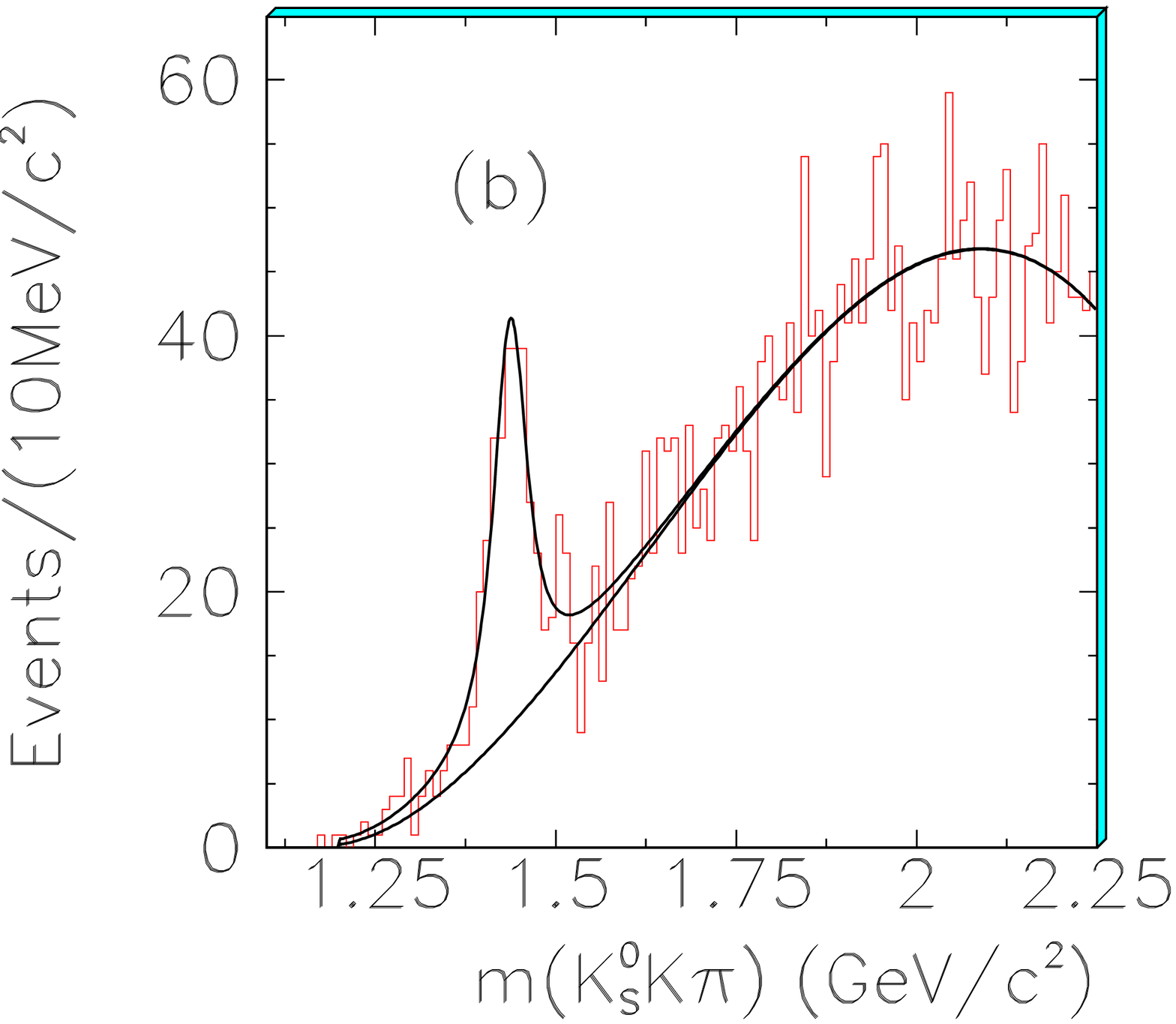}
\includegraphics[width=0.43\textwidth]{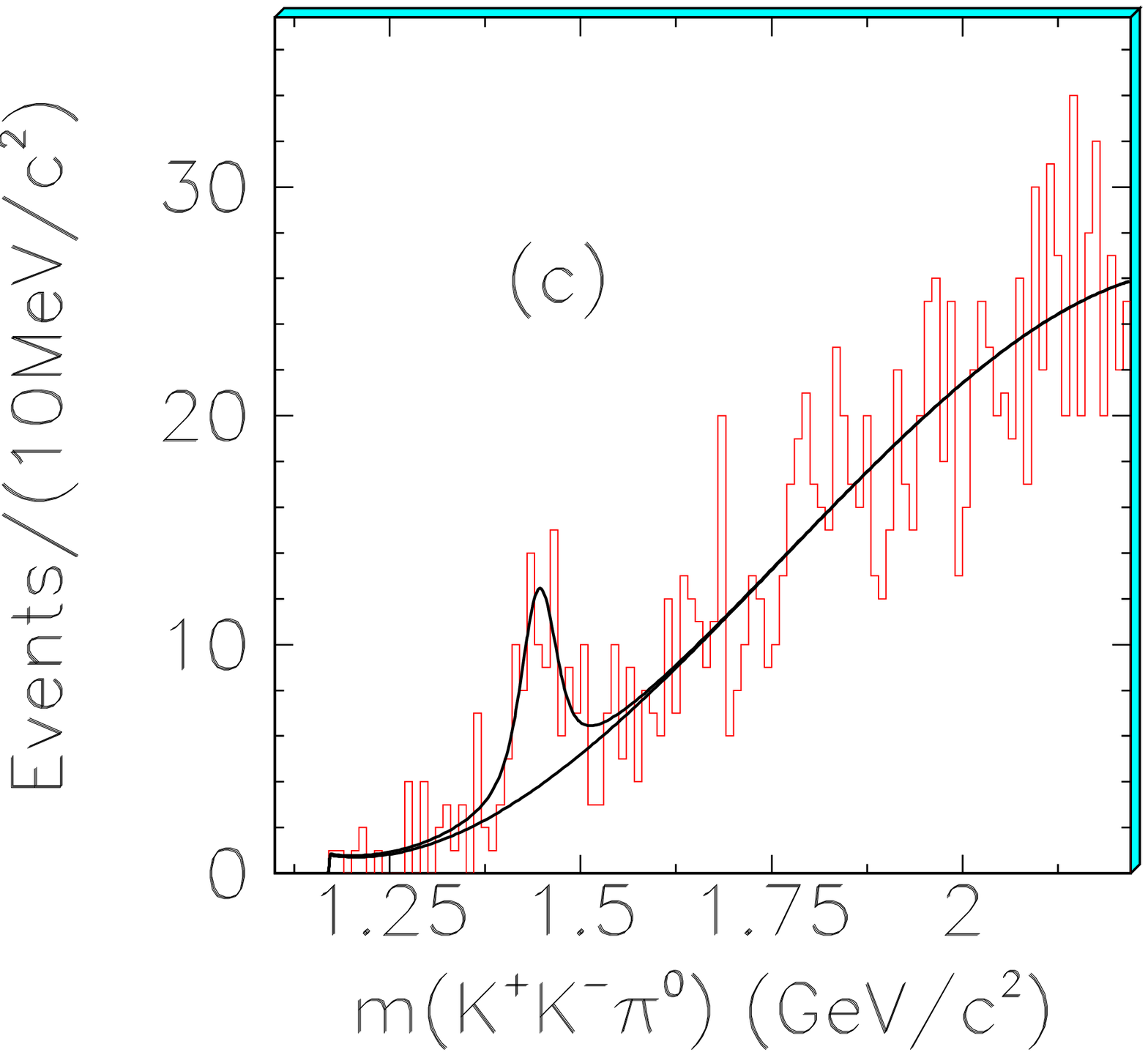}
\caption{The $K\bar{K}\pi$ invariant mass distribution for $J/\psi
\rightarrow \omega K^{0}_{S}K^{\pm}\pi^{\mp}$ (b) and $\omega
K^{+}K^{-}\pi^{0}$ (c) candidate events.  The curves are the best
fit. } \label{fig:w-x1440-recoiling}
\end{figure*}

\begin{figure*}[htbp]
  \centering
\includegraphics[width=0.43\textwidth]{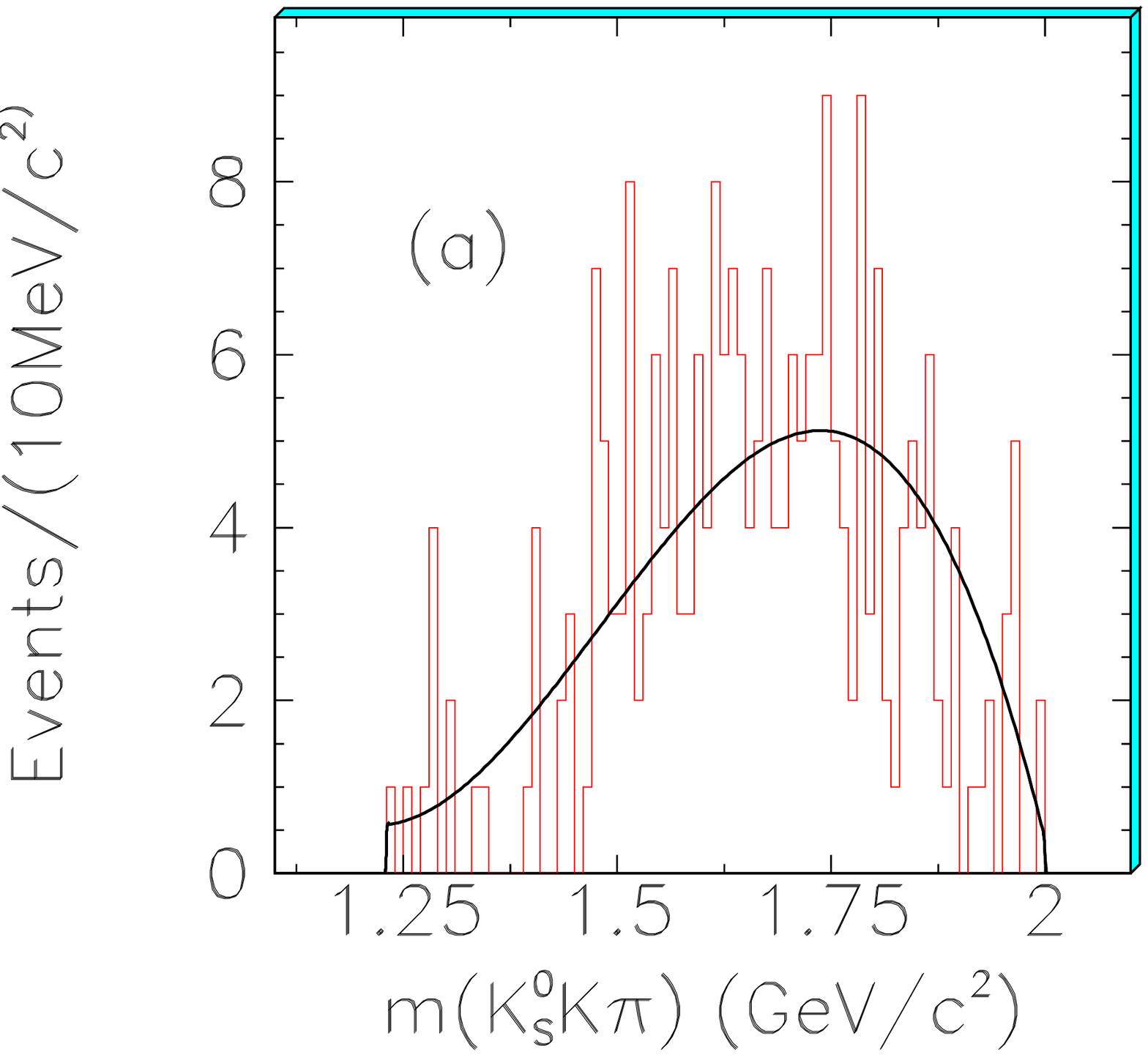}
\includegraphics[width=0.43\textwidth]{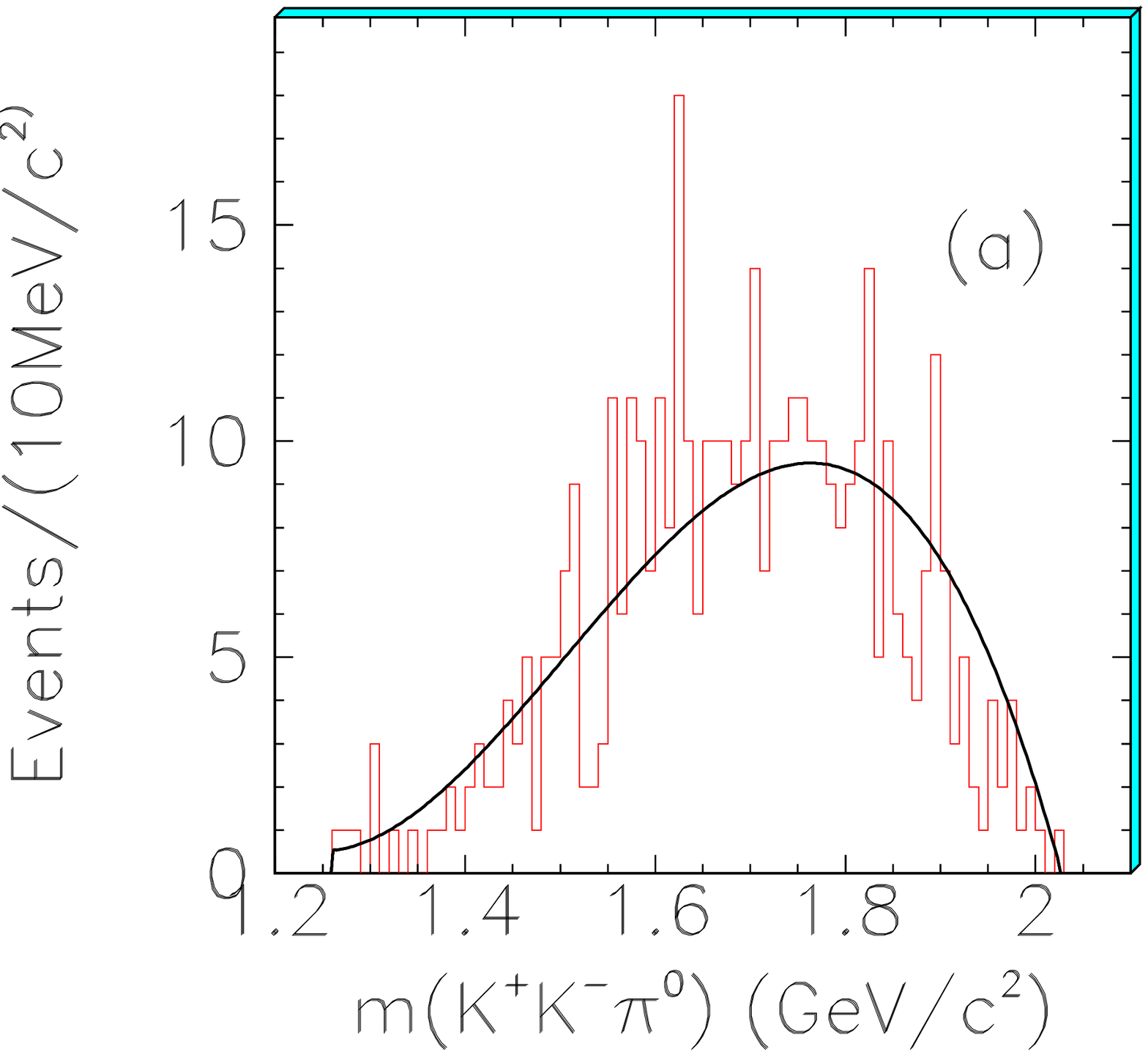}
\caption{The $K^{0}_{S} K^{\pm} \pi^{\mp}$  (left) and $\kk\pi^0$
(right) invariant mass recoiling against the $\phi$ in $\jpsi\to
\phi K\bar{K}\pi$ mode.} \label{fig:x1440-phikksp}
\end{figure*}

Figures~\ref{fig:w-x1440-recoiling} and~\ref{fig:x1440-phikksp} show
the $K^{0}_{S}K^{\pm}\pi^{\mp}$ and $K^{+} K^{-}\pi^{0}$ invariant
mass spectra after $\omega$ selection ($|m_{\pi^{+}\pi^{-}\gamma
\gamma}-m_{\omega}|<0.04$ GeV/c$^{2}$) or $\phi$ signal selection
($|m_{K^{+}K^{-}}-m_{\phi}|<0.015$~GeV/$c^{2}$). Clear $X(1440)$
signal is observed recoiling against the $\omega$, and there is no
significant signal recoiling against a $\phi$.

The $K^{0}_{S}K^{\pm}\pi^{\mp}$ invariant mass distribution in
$\jpsi\to \omega K^{0}_{S}K^{\pm}\pi^{\mp}$
(Fig.~\ref{fig:w-x1440-recoiling}(b)) is fitted with a BW function
convoluted with a Gaussian mass resolution function
($\sigma=7.44$~MeV/$c^{2}$) to represent the $X(1440)$ signal and a
third-order polynomial background function. The mass and width
obtained from the fit are $M=1437.6\pm 3.2$~MeV/$c^{2}$ and
$\Gamma=48.9 \pm 9.0$~MeV/$c^{2}$, and the fit yields $249\pm 35$
events. Using the efficiency of $1.45\%$ determined from a uniform
phase space MC simulation, we obtain the branching fraction to be
$\BR(J/\psi \rightarrow \omega X(1440))\cdot \BR( X(1440)
\rightarrow K^{0}_{S} K^{+} \pi^{-}+c.c.) = (4.86\pm 0.69\pm 0.81)
\times 10^{-4}$, where the first error is statistical and the second
one systematic.

For $\jpsi\to \omega K^{+} K^{-}\pi^{0}$ mode, by fitting the $K^{+}
K^{-}\pi^{0}$ mass spectrum in Fig.~\ref{fig:w-x1440-recoiling}(c)
with same functions, we obtain the mass and width of $M=1445.9\pm
5.7$~MeV/$c^{2}$ and $\Gamma=34.2\pm 18.5$~MeV/$c^{2}$, and the
number of events from the fit is $62\pm 18$. The efficiency is
determined to be $0.64\%$ from a phase space MC simulation, and the
branching fraction is $\BR(J/\psi \rightarrow \omega X(1440)) \cdot
\BR(X(1440) \rightarrow K^{+} K^{-} \pi^{0}) = (1.92\pm 0.57\pm
0.38) \times 10^{-4}$, in good agreement with the isospin symmetry
expectation from $\jpsi\to \omega K^{0}_{S}K^{\pm}\pi^{\mp}$ mode.

The distribution of $K^{0}_{S} K^{\pm} \pi^{\mp}$ and $K^{+} K^{-}
\pi^{0}$ invariant mass spectra recoiling against the $\phi$ signal
are shown in Fig.~\ref{fig:x1440-phikksp}, and there is no evidence
for $X(1440)$. The upper limits on the branching fractions at the
$90\%$ C.L. are $\BR(J/\psi\rightarrow \phi X(1440) \rightarrow \phi
K^{0}_{S}K^{+}\pi^{-}+c.c.)<1.93 \times 10^{-5}$ and $\BR( J/\psi
\rightarrow \phi X(1440) \rightarrow \phi K^{+}K^{-}\pi^{0}) < 1.71
\times 10^{-5}$.

In conclusion, the mass and width of the $X(1440)$ are measured,
which are in agreement with previous measurements; the branching
fractions we measured are also in agreement with the DM2 and
MARK-III results. The significant signal in $\jpsi\to \omega
K\bar{K}\pi$ mode and the missing signal in $\jpsi\to \phi X$ mode
may indicate the $s\bar{s}$ component in the $X(1440)$ is not
significant.
\section{New observations in $\jpsi$ and $\psp$ decays}

\subsection{\boldmath $\psp$ radiative decays}

Besides conventional meson and baryon states, QCD also predicts a
rich spectrum of glueballs, hybrids, and multi-quark states in the
1.0 to 2.5~$\hbox{GeV}/c^2$ mass region. Therefore, searches for the
evidence of these exotic states play an important role in testing
QCD. The radiative decays of $\psp$ to hadrons are expected to
contribute about 1\% to the total $\psp$ decay
width~\cite{PRD_wangp}. However, the measured channels only sum up
to about 0.05\%~\cite{PDG}.

\begin{figure}
  \centering
\includegraphics[width=0.48\textwidth,height=0.45\textheight]{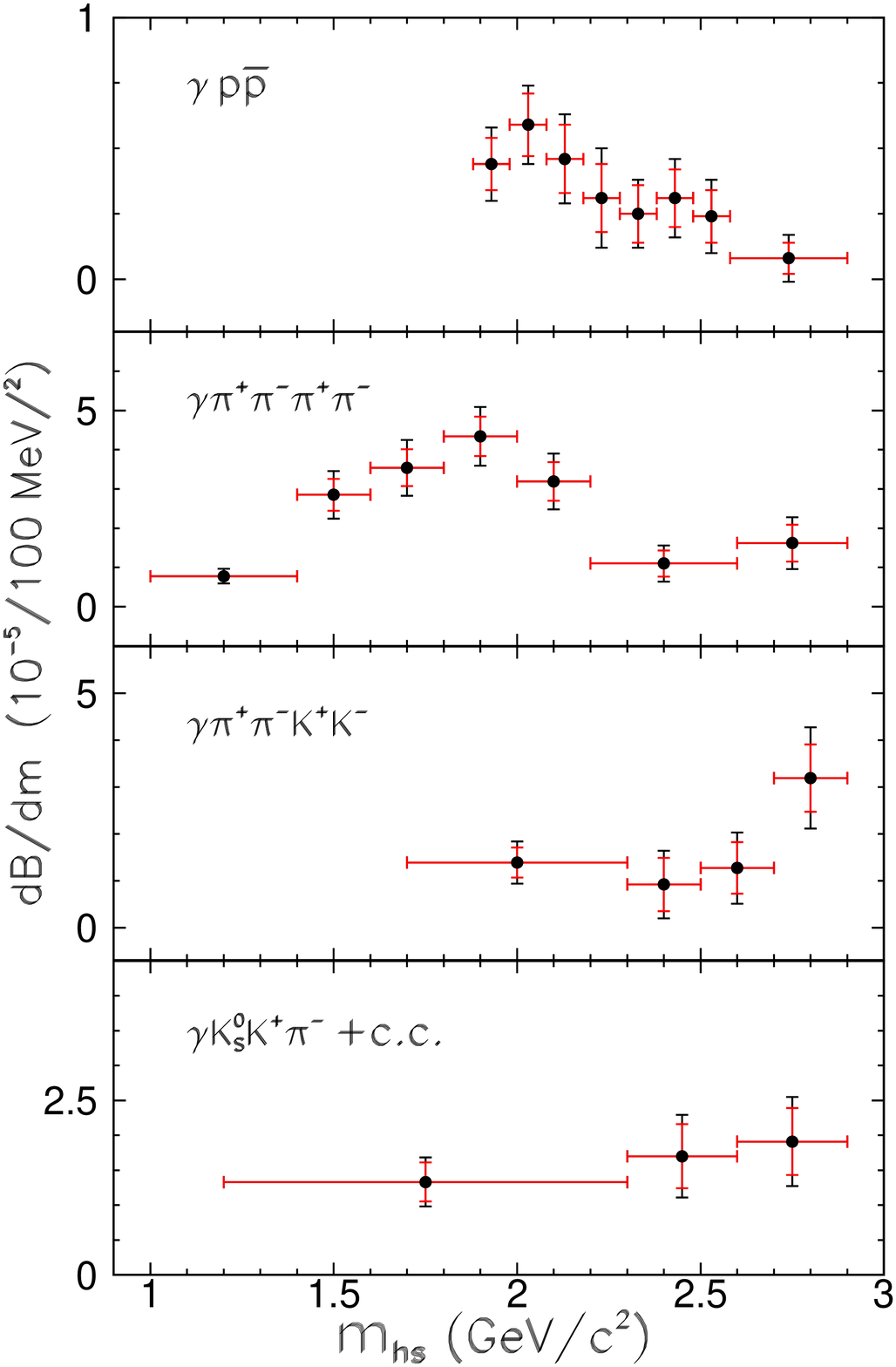}
\caption{ \label{difbr} Differential branching fractions for
$\psp\to \gamma\ppb$, $\gamma 2(\pipi)$, $\gamma K^+ K^- \pipi$, and
$\gamma \kskp$. Here $m_{hs}$ is the invariant mass of the hadrons.
For each point, the smaller longitudinal error is the statistical
error, while the bigger one is the total error. }
\end{figure}

We measured the decays of $\psp$ into $\gamma\ppb$, $\gamma
2(\pipi)$, $\gamma \kskp$, $\gamma K^+ K^- \pipi$, $\gamma
K^{*0}K^-\pi^+ +c.c.$, $\gamma K^{*0}\bar K^{*0}$,
$\gamma\pipi\ppb$, $\gamma 2(\kk)$, $\gamma 3(\pipi)$, and $\gamma
2(\pi^+\pi^-)K^+K^-$, with the invariant mass of the hadrons
($m_{hs}$) less than 2.9~$\hbox{GeV}/c^2$ for each decay
mode~\cite{bes2rad}. The differential branching fractions are shown
in Fig.~\ref{difbr}. The branching fractions below
$m_{hs}<2.9$~$\textrm{GeV}/c^2$ are given in Table~\ref{Tot-nev},
which sum up to $0.26\%$ of the total $\psp$ decay width. We also
analyzed $\psp\to \gamma\pipi$ and $\gamma\kk$ modes to study the
resonances in $\pipi$ and $\kk$ invariant mass spectrum. Significant
signals for $f_2(1270)$ and $f_0(1710)$ were observed, but the low
statistics prevent us from drawing solid conclusion on the other
resonances~\cite{agnes}.

\begin{table}
\caption{\label{Tot-nev} Branching fractions for $\psp\to\gamma
+hadrons$ with $m_{hs}<2.9$ $\hbox{GeV}/c^2$, where the upper limits
are determined at the 90\% C.L.}
\begin{center}
\begin{tabular}{ll} \hline \hline
Mode & $\BR(\times 10^{-5})$\\\hline
$\gamma p\bar{p}$ & 2.9$\pm$0.4$\pm$0.4 \\
$\gamma 2(\pi^+\pi^-)$ & 39.6$\pm$2.8$\pm$5.0\\
$\gamma K^0_S K^+\pi^-+c.c.$  & 25.6$\pm$3.6$\pm$3.6 \\
$\gamma K^+ K^-\pi^+\pi^-$ & 19.1$\pm$2.7$\pm$4.3 \\
$\gamma K^{*0} K^+\pi^-+c.c.$& 37.0$\pm$6.1$\pm$7.2\\
$\gamma K^{*0}\bar K^{*0}$&$24.0\pm 4.5\pm 5.0$\\
$\gamma \pipi\ppb$& 2.8$\pm$1.2$\pm$0.7 \\
$\gamma \kk\kk$ &  $<4$\\
$\gamma3(\pipi)$&  $<17$\\
$\gamma2(\pi^+\pi^-)K^+K^-$& $<22$ \\
\hline \hline
\end {tabular}
\end{center}
\end{table}
\subsection{$\jpsi,~\psp\to n\ks\ldb+c.c.,~\ld\ldb\pi^0,~\ld\ldb\eta$}
The $X(2075)$ was first reported by BESII near the threshold of the
invariant mass spectrum of $p\ldb$ in $\jpsi\to pK^-\ldb$ decays.
The mass, width, and product branching fraction of this enhancement
are $M = 2075 \pm 12~({\rm stat.}) \pm 5~({\rm syst.})$ MeV/$c^2$,
$\Gamma = 90 \pm 35~({\rm stat.}) \pm 9~({\rm syst.})$
MeV/$c^2$~\cite{pkl}, and $B(J/\psi \to K^- X)B(X \to p
\bar{\Lambda}+c.c.) = (5.9 \pm 1.4 \pm 2.0) \times 10^{-5}$,
respectively.  The study of the isospin conjugate channel $\jpsi\to
n K^0_S \bar{\Lambda}$ is therefore important not only in exploring
new decay modes of $J/\psi$ but also in understanding the $X(2075)$.

\begin{figure}[htbp]
\begin{center}
\hspace*{-0.cm} \epsfysize=7cm \epsffile{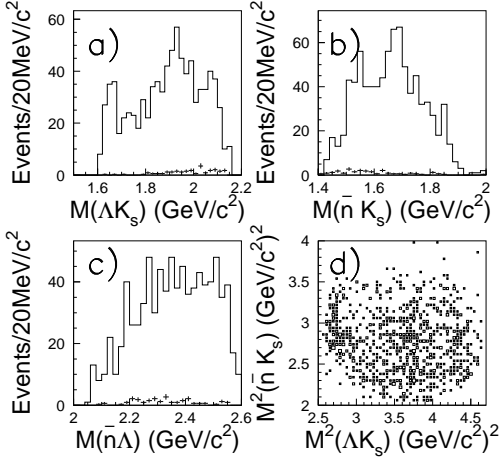}
\end{center}
%\centerline{\hbox{\psfig{file=Nstar.eps,width=9cm,height=9cm}}}
\caption{The invariant mass spectra of (a) ${\Lambda K_S^0}$, (b) ${\bar n K_S^0}$,
  and (c) $\bar {n} \Lambda$, as well as (d)
  the Dalitz plot for candidate events after all selection criteria.
  The crosses show the sideband backgrounds.}
\label{Nstar}
\end{figure}

 The
invariant mass spectra of ${\Lambda K_S^0}$, ${n K_S^0}$, and ${\bar
{\Lambda} n (\Lambda \bar n)}$, as well as the Dalitz plot for all
selection requirements are shown in Fig.~\ref{Nstar}.  In the
$\Lambda K_S^0$ invariant mass spectrum, an enhancement near
$\Lambda K_S^0$ threshold is evident, as is found in the $\Lambda K$
mass spectrum in $J/\psi \to p K^- \bar{\Lambda}$~\cite{pkl2}. If
the enhancement is fitted with an acceptance weighted S-wave
Breit-Wigner function and a function $f_{bg}(\delta)$ describing the
phase space ``background'' contribution, the fit leads to
M=1.648$\pm$0.006GeV/$c^2$ and $\Gamma=61\pm21$MeV/$c^2$,
respectively. Here the errors are only statistical. The systematic
uncertainties are not included since more accurate measurements of
the mass and width should come from a full PWA involving
interferences among $N^*$ and $\Lambda^*$ states. The fitted mass
and width are consistent with those obtained from a partial wave
analysis of $J/\psi \to p K^- \bar{\Lambda}$~\cite{pkl2}. The
$X(2075)$ signal which was seen in the $p\bar{\Lambda}$ invariant
mass spectrum in $\jpsi\to p K^- \bar{\Lambda}$ is not significant
here.  Using a Bayesian approach~\cite{Bayesian} and fixing the mass
and width of $X(2075)$ to 2075 MeV/$c^2$ and 90 MeV/$c^2$
respectively, the upper limit on the number of events observed
$N_{obs}^{UL}$ is 54 events at the 90\% C.L.

\vspace{0.5cm}

 The decays of $\jpsi$ and $\psp$ to $n K^0_S\bar\Lambda + c.c.$  are
observed for the first time, and their branching fractions are:
\begin{center}
$B(\jpsi\to n K^0_S\bar\Lambda+c.c.)=(6.46\pm0.20\pm1.07)\times 10^{-4}$,\\
$B(\jpsi\to n K^0_S\bar\Lambda)=(3.09\pm0.14\pm0.58)\times 10^{-4}$,\\
$B(\jpsi\to \bar n K^0_S\Lambda)=(3.39\pm0.15\pm0.48)\times 10^{-4}$,\\
$B(\psp\to n K^0_S\bar\Lambda+c.c.)=(0.81\pm0.11\pm0.14)\times 10^{-4}$.
\end{center}

The isospin violating process $\jpsi\to\ld\ldb\pi^0$ has been studied by
DM2~\cite{np3} and BESI~\cite{np4}, and its average branching fraction
is determined to be ${\cal B}(\jpsi\to\llp)=(2.2\pm 0.6)\times
10^{-4}$~\cite{np12}. However, the isospin conserving process
$\jpsi\ar\lle$ has not been reported, and there are no measurements
for $\llp$ and $\lle$ decays of $\psp$.

Table ~\ref{branresult} lists the results for $\jpsi$ and $\psp$ decay
into $\llp$ and $\lle$, as well as $\jpsi\ar\splb+c.c.$. We also list the total branching fraction
for the conjugate modes, where the common systematic errors have been taken out. Except for
$\jpsi\ar\llp$ and $\jpsi\ar\splb+c.c.$, the results are first
measurements. Interestingly, the result of $\jpsi\ar\llp$
presented here is much smaller than those of DM2 and
BESI~\cite{np3,np4}. In previous experiments, the large contaminations
from $\jpsi\ar\Sigma^0\pi^0\lmb+c.c.$ and
$\jpsi\ar\Sigma^+\pi^-\lmb+c.c.$ were not considered, resulting in a
large value of branching fraction for $\jpsi\ar\llp$. The small
branching fraction of $\jpsi\ar\llp$ and relatively large branching
fraction of $\jpsi\ar\lle$ measured here indicate that the isospin
violating decay in $\jpsi$ decays is suppressed while isospin
conserving decays are favored, which is consistent with
expectation.

\begin{table*}
\caption{Measured branching fractions or upper limits at 90\% confidence level (C.L.) for all the studied channels. Here, ${\cal B}(\Lambda\ar\pi^- p)=63.9\%$,
 ${\cal B}(\Sigma^+\ar\pi^0 p)=51.6\%$ and ${\cal
 B}(\eta\ar\GG)=39.4\%$ are taken from the PDG.}
\bcl
%\doublerulesep 2pt
\begin{tabular}{l|c|c|c}\hline\hline
Channels&Number of events&MC efficiency(\%)&Branching fraction ($\times 10^{-4}$)\\ \hline
$\jpsi\ar\llp$&$<11.2$&0.75&$<0.64$\\
$\jpsi\ar\lle$&$44\pm 10$&$1.8$&$2.62\pm 0.60\pm 0.44$\\ \hline
$\psp\ar\llp$&$<7.0$&2.5&$<0.49$\\
$\psp\ar\lle$&$<7.6$&2.9&$<1.2$\\ \hline
$\jpsi\ar\splb$&$335 \pm 22$&2.3&$7.70\pm 0.51\pm 0.83$ \\
$\jpsi\ar\sbpl$&$254 \pm 19$&1.8&$7.47\pm 0.56\pm 0.76$ \\ \hline
$\jpsi\ar\splb+c.c.$& & &$15.17\pm 0.76\pm 1.59$ \\  \hline
\end{tabular}
\label{branresult}
\ecl
\end{table*}

\section{Search for $\jpsi$ rare decays via $\psp\to\pi^+\pi^-\jpsi$}
Search for $\jpsi$ rare decays, e.g. $C$-parity violation or invisible decays, suffers from
removing the QED backgrounds from the direct annihilation of $e^+e^-$. Using the $\jpsi$ sample produced from
$\psp\to\pi^+\pi^-\jpsi$, the QED background can be strongly suppressed.
 The direct decay $\jpsi\to\GG$ was previously measured
to be $Br(\jpsi\to\GG)<5\times 10^{-4}$. BES studies the decay $\jpsi\to\GG$ using $\jpsi\to\pi^+\pi-\jpsi$, and the upper limit for
the branching ratio is measured to be $Br(\jpsi\to\GG)<2.2\times 10^{-5}$ at 90\% confidence level, which is about
20 times lower than previous measurements.

Invisible decays of quarkonium states offer a window into what may lie beyond the standard model. In standard model (SM),
the predicted branching fraction for $\jpsi\to\nu\nu$ is $Br(\jpsi\to\nu\nu)=4.54\times 10^{-7}\times Br(\jpsi\to e^+e^-)$
 with a small uncertainty (2\%-3\%). However, new physics beyond the SM might enhance the branching fraction of
 $\jpsi$ invisible decays. One possibility is the decay into light dark matter particles mediated by a new, electrically
 neutral spin-1 gauge boson $U$, which could significantly increase the invisible decay rate \cite{darkmatter}. It is of interest
 to search for such light invisible particle in collider experiments. Using $\psp\to \pi^+\pi^-\jpsi$ decays,  a
  search for the decay of the $\jpsi$ to invisible final states
  is performed. The $\jpsi$ peak in the distribution of
  masses recoiling against the $\pi^+\pi^-$ is used to tag $\jpsi$ invisible
  decays. No signal is found, and an upper limit at the 90\% confidence
  level is determined to be $1.2\times 10^{-2}$ for the ratio
  $\frac{{\cal B}(\jpsi\ar \mbox{invisible})}{{\cal B}(\jpsi\ar\mu^+\mu^-)}$.
  This is the first search for $\jpsi$ decays to invisible final
  states.
\section{\boldmath Summary}

Using the 58~M $\jpsi$ and 14~M $\psp$ event samples taken with the
BESII detector at the BEPC storage ring, BES experiment provided
many interesting results in charmonium decays, including the
observation of the $Y(2175)$, $\eta(2225)$, $X(1440)$, and many
$\psp$ radiative decays. The effort to search for rare decays,
 e.g. $J/\psi $ decays into $\gamma\gamma$ and invisible decays are also reported.
 These results shed light on the
understanding of role played by strong interactions in charmonium decays.

\begin{acknowledgments}
This work is supported in part by the National Natural Science Foundation of China under
contracts Nos. 10491300, 10225524, 10225525, 10425523,10625524,
10521003.
\end{acknowledgments}

\end{document}